\def\etal{{\frenchspacing\it et al.}}

\def\beq#1{\begin{equation}\label{#1}}
\def\eeq{\end{equation}}
\def\beqa#1{\begin{eqnarray}\label{#1}}
\def\eeqa{\end{eqnarray}}

\def\fun#1#2{\lower3.6pt\vbox{\baselineskip0pt\lineskip.9pt
        \ialign{$\mathsurround=0pt#1\hfill##\hfil$\crcr#2\crcr\sim\crcr}}}
        
\def\xi{{{\bf x}^b}}

\documentclass[twocolumn,aps,showpacs,showkeys,nofootinbib]{revtex4}
\usepackage{epsfig}

\newcommand{\be}{\begin{equation}}
\newcommand{\ee}{\end{equation}}
\newcommand{\ba}{\begin{eqnarray}}
\newcommand{\ea}{\end{eqnarray}}

\begin{document}
\input{epsf.sty}

\title{A Comparative Study of Dark Energy Constraints from Current Observational Data}
\author{Yun~Wang$^{1}$\footnote{email: wang@nhn.ou.edu}, 
Chia-Hsun~Chuang$^1$, \& Pia~Mukherjee$^2$}
\address{$^1$Homer L. Dodge Department of Physics \& Astronomy, Univ. of Oklahoma,
                 440 W Brooks St., Norman, OK 73019\\
         $^2$Sussex Astronomy Centre, Dept. of Physics \& Astronomy, University of Sussex,
		Falmer, Brighton, BN1 9QH, U.K.}

                 \today

\begin{abstract}

We examine how dark energy constraints from current observational data 
depend on the analysis methods used: the analysis of
Type Ia supernovae (SNe Ia), and that of galaxy clustering data.
We generalize the flux-averaging analysis method of SNe Ia to 
allow correlated errors of SNe Ia, in order to reduce the systematic bias 
due to weak lensing of SNe Ia. We find that flux-averaging leads to 
larger errors on dark energy and cosmological parameters if only SN Ia 
data are used. When SN Ia data (the latest compilation by the SNLS team)
are combined with WMAP 7 year results (in terms of our Gaussian fits to 
the probability distributions of the CMB shift parameters), the latest 
Hubble constant ($H_0$) measurement using the Hubble Space Telescope (HST), 
and gamma ray burst (GRB) data, flux-averaging of SNe Ia increases the concordance  
with other data, and leads to significantly tighter constraints on 
the dark energy density at $z=1$, and the cosmic curvature $\Omega_k$.
The galaxy clustering measurements of $H(z=0.35)r_s(z_d)$ and 
$r_s(z_d)/D_A(z=0.35)$ (where $H(z)$ is the Hubble parameter, 
$D_A(z)$ is the angular diameter distance, and $r_s(z_d)$ is the
sound horizon at the drag epoch) by Chuang \& Wang (2011) are consistent with
SN Ia data, given the same pirors (CMB+$H_0$+GRB), and lead to significantly 
improved dark energy constraints when combined. Current 
data are fully consistent with a cosmological constant and a flat universe.

\end{abstract}

%98.80.Es Observational cosmology (including Hubble constant, 
%distance scale, cosmological constant, early Universe, etc)
%98.80.-k Cosmology 
%98.80.Jk Mathematical and relativistic aspects of cosmology)


\maketitle

\section{Introduction}

Solving the mystery of cosmic acceleration \cite{Riess98,Perl99}
is one of the most important challenges in cosmology 
today. Current observational data are not sufficient for
differentiating two likely explanations for the observed
cosmic acceleration: dark energy, and the modification
of general relativity. For recent reviews, see 
\cite{Copeland06,Ruiz07,Ratra07,Frieman08,Caldwell09,Uzan09,Wang10,Li11}.

There are a number of powerful direct probes of dark energy: Type 
Ia supernovae (SNe Ia) \cite{Riess98,Perl99}; galaxy clustering (GC), 
especially the baryon acoustic oscillations (BAO) \cite{BG03,SE03}; 
and weak lensing of galaxies \cite{Hu02,Jain03}. These
methods are complementary to each other, and each method has its own
systematic uncertainties. Other data, e.g., gamma ray bursts
\cite{Amati02,Bloom03,Schaefer03}, can help strengthen the dark 
energy constraints.

The cosmic microwave background (CMB) anisotropy data provide 
strong priors on cosmological parameters (see, e.g., \cite{Komatsu11}).
Direct measurements on the Hubble constant (see, e.g., \cite{Riess11})
also help break the degeneracy amongst the dark energy and cosmological 
parameters.

Much progress has been made since the discovery of cosmic
acceleration in 1998. However, current data are still rather
limited in constraining dark energy. In particular, different
analysis methods of the same data could lead to significantly
different constraints on dark energy. This is probably due to 
the fact that different methods have different residual systematic
biases, and they often assume different priors on cosmological 
parameters as well.

In this paper we examine these issues by studying how dark energy 
constraints from current observational data depend on the analysis 
methods used. In particular, we explore the impact on the overall
dark energy constraints of 
(1) using flux-averaging to minimize the weak lensing bias in the 
analysis of SNe Ia; 
(2) using radial and transverse distance constraints derived
from GC data without assuming CMB priors.

We describe our method in Sec.II, present our results in Sec.III,
and conclude in Sec.IV.

\section{Method}
\label{sec:method}

We only use methods that give geometric constraints 
on dark energy. The constraints on the growth rate of
cosmic large scale structure are degenerate with the
geometric constraints (see, e.g., \cite{Wang08a,Simpson10});
current data do not allow the determination of both
without strong assumptions, e.g., assuming that general 
relativity is not modified. 

Geometric constraints on dark energy are derived from
the measurement of distances.
The comoving distance to an object at redshift $z$ is given by:
\ba
\label{eq:r(z)}
 & &r(z)=cH_0^{-1}\, |\Omega_k|^{-1/2} {\rm sinn}[|\Omega_k|^{1/2}\, \Gamma(z)],\\
 & &\Gamma(z)=\int_0^z\frac{dz'}{E(z')}, \hskip 1cm E(z)=H(z)/H_0 \nonumber
\ea
where ${\rm sinn}(x)=\sin(x)$, $x$, $\sinh(x)$ for 
$\Omega_k<0$, $\Omega_k=0$, and $\Omega_k>0$ respectively;
and the expansion rate of the universe $H(z)$ (i.e.,
the Hubble parameter) is given by
\ba
\label{eq:H(z)}
&&H^2(z)  \equiv  \left(\frac{\dot{a}}{a}\right)^2 \\
 &= &H_0^2 \left[ \Omega_m (1+z)^3 +\Omega_r (1+z)^4 +\Omega_k (1+z)^2 
+ \Omega_X X(z) \right],\nonumber
\ea
where $\Omega_m+\Omega_r+\Omega_k+\Omega_X=1$, and
the dark energy density function $X(z)$ is defined as
\be
X(z) \equiv \frac{\rho_X(z)}{\rho_X(0)}.
\ee
Note that $\Omega_r \ll \Omega_m$, thus the $\Omega_r$ term
is usually omitted in dark energy studies, since dark energy
should only be important at late times.

\subsection{Analysis of SN Ia Data}

SN Ia data give measurements of the luminosity distance 
$d_L(z)$ through that of the distance modulus
of each SN:
\be
\label{eq:m-M}
\mu_0 \equiv m-M= 5 \log\left[\frac{d_L(z)}{\mathrm{Mpc}}\right]+25,
\ee
where $m$ and $M$ represent the apparent and absolute magnitude
of a SN. The luminosity distance $d_L(z)=(1+z)\, r(z)$, with the comoving
distance $r(z)$ given by Eq.(\ref{eq:r(z)}). 

We use the compilation of SN Ia data by Conley et al. (2011) \cite{Conley11}, 
which include the SNe Ia from the first three years of the Supernova Legacy 
Survey (SNLS3), the largest homogeneous SN Ia data set.
We compare two methods for using these SNe in constraining
dark energy: with flux-averaging of SNe Ia, and without.

\subsubsection{SN Ia Data}

For a set of 472 SNe Ia, Conley et al. (2011) \cite{Conley11} give 
the apparent $B$ magnitude, $m_B$, and the covariance matrix for
$\Delta m \equiv m_B-m_{\rm mod}$, with
\be
m_{\rm mod}=5 \log_{10}{\cal D}_L(z|\mbox{\bf s})
- \alpha (s-1) +\beta {\cal C} + {\cal M},
\ee
where ${\cal D}_L(z|\mbox{\bf s})$ is the luminosity distance
multiplied by $H_0$
for a given set of cosmological parameters $\{ {\bf s} \}$,
$s$ is the stretch measure of the SN light curve shape, and
${\cal C}$ is the color measure for the SN.
${\cal M}$ is a nuisance parameter representing some combination
of the absolute magnitude of a fiducial SN Ia, $M$, and the 
Hubble constant $H_0$.
%mu0pa(j)=5.*log10((1.+zhel(j))*rcov0)-alpha*(str(j)-1.0)+beta*color(j)+Msn
Since the time dilation part of the observed luminosity distance depends 
on the total redshift $z_{\rm hel}$ (special relativistic plus cosmological),
we have \cite{Hui06}
\be
{\cal D}_L(z|\mbox{\bf s})\equiv c^{-1}H_0 (1+z_{\rm hel}) r(z|\mbox{\bf s}),
\ee
where $z$ and $z_{\rm hel}$ are the CMB restframe and heliocentric redshifts
of the SN. 

For a set of $N$ SNe with correlated errors, we have \cite{Conley11}
\be
\label{eq:chi2_SN}
\chi^2=\Delta \mbox{\bf m}^T \cdot \mbox{\bf C}^{-1} \cdot \Delta\mbox{\bf m}
\ee
where $\Delta \bf m$ is a vector with $N$ components, and
$\mbox{\bf C}$ is the $N\times N$ covariance matrix of the SNe Ia.

Note that $\Delta m$ is equivalent to $\Delta \mu_0$, since 
\be
\Delta m \equiv m_B-m_{\rm mod}
= \left[m_B+\alpha (s-1) -\beta {\cal C}\right] - {\cal M}.
\ee

The total covariance matrix is \cite{Conley11}
\be
\mbox{\bf C}=\mbox{\bf D}_{\rm stat}+\mbox{\bf C}_{\rm stat}
+\mbox{\bf C}_{\rm sys},
\ee
with the diagonal part of the statistical uncertainty given by \cite{Conley11}
\ba
\mbox{\bf D}_{\rm stat,ii}&=&\sigma^2_{m_B,i}+\sigma^2_{\rm int}
+ \sigma^2_{\rm lensing}+ \sigma^2_{{\rm host}\,{\rm correction}} \nonumber\\
&& + \left[\frac{5(1+z_i)}{z_i(1+z_i/2)\ln 10}\right]^2 \sigma^2_{z,i} 
 +\alpha^2 \sigma^2_{s,i}+\beta^2 \sigma^2_{{\cal C},i} \nonumber\\
&& + 2 \alpha C_{m_B s,i} - 2 \beta C_{m_B {\cal C},i}
-2\alpha\beta C_{s {\cal C},i},
\ea
where $C_{m_B s,i}$, $C_{m_B {\cal C},i}$, and $C_{s {\cal C},i}$
are the covariances between $m_B$, $s$, and ${\cal C}$ for the $i$-th SN. 
Note also that $\sigma^2_{z,i}$ includes a
peculiar velocity residual of 0.0005 (i.e., 150$\,$km/s) added 
in quadrature \cite{Conley11}.

The statistical and systematic covariance matrices, 
$\mbox{\bf C}_{\rm stat}$ and $\mbox{\bf C}_{\rm sys}$,
are generally not diagonal \cite{Conley11}, and are given in the
form:
\be
\mbox{\bf C}_{\rm stat}+\mbox{\bf C}_{\rm sys}
=V_0+\alpha^2 V_a + \beta^2 V_b + 2 \alpha V_{0a}
-2 \beta V_{0b} - 2 \alpha\beta V_{ab}.
\ee
where $V_0$, $V_{a}$, $V_{b}$, $V_{0a}$, $V_{0b}$, and
$V_{ab}$ are matrices given by the SNLS data archive
at https://tspace.library.utoronto.ca/handle/1807/24512/.
$\mbox{\bf C}_{\rm stat}$ includes the uncertainty in
the SN model. $\mbox{\bf C}_{\rm sys}$ includes the
uncertainty in the zero point. Note that  $\mbox{\bf C}_{\rm stat}$
and $\mbox{\bf C}_{\rm sys}$ do not depend on ${\cal M}$, 
since the relative distance moduli are independent of the value 
of ${\cal M}$ \cite{Conley11}.

We refer the reader to Conley et al. (2011) \cite{Conley11}
for a detailed discussion of the origins of the statistical
and systematic errors. As an example, we note that the correlation 
of errors on different SNe arises from a statistical uncertainty 
in the zero point of one passband, e.g., $r_M$.
This directly affects all SNe with $r_M$ measurements
due to K-corrections (restframe $B$ to $r_M$), and
indirectly affects even the SNe without $r_M$ measurements 
through the empirical SN models by changing
the templates and the measured color-luminosity relationship.

\subsubsection{The Recipe for Flux-Averaging of SNe Ia}

Because we live in a lumpy universe, SN Ia observations
can be misinterpreted if the effect of gravitational lensing 
is not properly accounted for (see, e.g., \cite{Clarkson11}).
Flux-averaging of SNe Ia was proposed to reduce the effect of the weak 
lensing of SNe Ia on cosmological parameter estimation \cite{Wang00}. 
The basic idea is that because of flux conservation in gravitational
lensing, the average magnification of a large number of SNe Ia at
the same redshift should be unity. Thus averaging the observed flux from
a large number of SNe Ia at the same redshift can recover the unlensed
brightness of the SNe Ia at that redshift.

Wang \& Mukherjee (2004) \cite{WangPia04} and Wang (2005) 
\citep{Wang05} developed a consistent framework for flux-averaging SNe Ia. 
Wang \& Mukherjee (2004) \cite{WangPia04} gave the recipe for flux-averaging 
SNe Ia in the absence of correlated errors.
Here we modify and generalize the recipe to allow correlated errors of SNe Ia.

As described in \cite{Wang00}, the fluxes of SNe Ia in a redshift bin
should only be averaged {\it after} removing their redshift dependence,
which is a model-dependent process.
For $\chi^2$ statistics using MCMC or a grid of parameters, 
here are the steps in flux-averaging:

(1) Convert the distance modulus of SNe Ia into 
``fluxes'',
\be
\label{eq:flux}
F(z_l) \equiv 10^{-(\mu_0^{\rm data}(z_l)-25)/2.5} =  
\left( \frac{d_L^{\rm data}(z_l)} {\mbox{Mpc}} \right)^{-2}.
\ee

(2) For a given set of cosmological parameters $\{ {\bf s} \}$,
obtain ``absolute luminosities'', \{${\cal L}(z_l)$\}, by
removing the redshift dependence of the ``fluxes'', i.e.,
\be
\label{eq:lum}
{\cal L}(z_l) \equiv d_L^2(z_l |{\bf s})\,F(z_l).
\ee

(3) Flux-average the ``absolute luminosities'' \{${\cal L}^i_l$\} 
in each redshift bin $i$ to obtain $\left\{\overline{\cal L}^i\right\}$:
\be 
 \overline{\cal L}^i = \frac{1}{N_i}
 \sum_{l=1}^{N_i} {\cal L}^i_l(z^{(i)}_l),
 \hskip 1cm
 \overline{z_i} = \frac{1}{N_i}
 \sum_{l=1}^{N_i} z^{(i)}_l. 
\ee

(4) Place $\overline{\cal L}^i$ at the mean redshift $\overline{z}_i$ of
the $i$-th redshift bin, now the binned flux is
\be
\overline{F}(\overline{z}_i) = \overline{\cal L}^i /
d_L^2(\overline{z}_i|\mbox{\bf s}).
\ee

(5) Compute the covariance matrix of $\overline{F}(\overline{z}_i)$
and $\overline{F}(\overline{z}_j)$:
\ba
&& \mbox{Cov}\left[\overline{F}(\overline{z}_i),\overline{F}(\overline{z}_j)\right] \\
&=&\frac{1}{N_i N_j}\left[\frac{\ln 10 /2.5}
{d_L(\overline{z}_i|\mbox{\bf s})d_L(\overline{z}_j|\mbox{\bf s})}
\right]^2 \cdot \nonumber\\
&& \sum_{l=1}^{N_i} \sum_{m=1}^{N_j} {\cal L}(z_l^{(i)})
{\cal L}(z_m^{(j)}) \langle \Delta \mu_0^{\rm data}(z_l^{(i)})\Delta 
\mu_0^{\rm data}(z_m^{(j)})
\rangle \nonumber 
\ea
where $\langle \Delta \mu_0^{\rm data}(z_l^{(i)})\Delta \mu_0^{\rm data}(z_m^{(j)})\rangle $
is the covariance of the measured distance moduli of the $l$-th SN Ia
in the $i$-th redshift bin, and the $m$-th SN Ia in the $j$-th
redshift bin. ${\cal L}(z)$ is defined by Eqs.(\ref{eq:flux}) and (\ref{eq:lum}).

(6) For the flux-averaged data, $\left\{\overline{F}(\overline{z}_i)\right\}$, 
compute
\be
\label{eq:chi2_SN_fluxavg}
\chi^2 = \sum_{ij} \Delta\overline{F}(\overline{z}_i) \,
\mbox{Cov}^{-1}\left[\overline{F}(\overline{z}_i),\overline{F}(\overline{z}_j)
\right] \,\Delta\overline{F}(\overline{z}_j)
\ee
where
\be
\Delta\overline{F}(\overline{z}_i) \equiv 
\overline{F}(\overline{z}_i) - F^p(\overline{z}_i|\mbox{\bf s}),
\ee
with $F^p(\overline{z}_i|\mbox{\bf s})=
\left( d_L(z|\mbox{\bf s}) /\mbox{Mpc} \right)^{-2}$.

For the sample of SNe we use in this study, we  
flux-averaged the SNe with $dz=0.07$, to ensure that
all redshift bins contain at least one SN.
Our SN flux-averaging code is available 
at http://www.nhn.ou.edu/$\sim$wang/SNcode/.

\subsection{Galaxy Clustering Data}
\label{sec:GC}

For GC data, we use the distance measurements
from the SDSS DR7 data. We compare two sets of distance
measurements from SDSS DR7: (1) ``CW2'': the measurement of
$H(z)r_s(z_d)$ and $r_s(z_d)/D_A(z)$ (where $H(z)$ is the Hubble parameter, 
$D_A(z)$ is the angular diameter distance, and $r_s(z_d)$ is the
sound horizon at the drag epoch) by Chuang \& Wang (2011)
\cite{CW11} from the two-dimensional two-point correlation
function of SDSS DR7 Luminous Red Galaxies (LRGs);
(2) ``WP2'': the measurement of $r_s(z_d)/D_V(z=0.2)$ and
$r_s(z_d)/D_V(z=0.35)$ (where $D_V(z)$ is the spherically-averaged
distance) by Percival et al. (2010) \cite{Percival10}
from the spherically-averaged power spectrum of combined data 
of SDSS DR7 LRG and main galaxy samples and 2-degree Field Galaxy 
Redshift Survey (2dFGRS).

Using the two-dimensional two-point correlation function of SDSS DR7
in the scale range of 40-120$\,$Mpc/$h$, 
Chuang \& Wang (2011) \cite{CW11} found that
\ba
H(z=0.35)r_s(z_d)&=&13020\pm  530\,{\rm km/s} \,\,(4\%) \nonumber \\
r_s(z_d)/D_A(z=0.35)&=& 0.1518\pm  0.0062\,\, (4\%) \nonumber\\
r&=&-0.0584,
\label{eq:CW2}
\ea
where $r$ is the normalized correlation coefficient between
$H(z=0.35)r_s(z_d)$ and $r_s(z_d)/D_A(z=0.35)$, and 
$r_s(z_d)$ is the sound horizon at the drag epoch (given by
Eqs.(\ref{eq:rs}) and (\ref{eq:zd}). The inverse covariance
matrix of $H(z=0.35)r_s(z_d)$ and $r_s(z_d)/D_A(z=0.35)$ is
\be
C_{\rm GC,CW2}^{-1}=\left( \begin{array}{cc}
   0.35722E-05  & 0.17833E-01\\
   0.17833E-01  & 0.26104E+05
   \end{array}
\right)
\label{eq:CW2_cov}
\ee

Spherically-averaged data give a measurement of
$d_z \equiv r_s(z_d) /D_V(z)$, with
\be
D_V(z) \equiv \left[ \frac{ r(z)^2\, c z }{H(z)}\right]^{1/3}.
\ee
Chuang \& Wang (2011) found that 
\be
\label{eq:CW1}
d_{0.35}^{CW} \equiv r_s(z_d)/D_V(z=0.35)=0.1161\pm   0.0034\,\, (2.9\%)
\ee 
from the spherically-averaged correlation function of SDSS LRGs.

Using the spherically-averaged power spectrum of combined data 
of SDSS DR7 LRG and main galaxy samples and 2dFGRS,
Percival et al. (2010) \cite{Percival10} found that
\ba
d_{0.2}^{WP}&\equiv& r_s(z_d)/D_V(z=0.2)=0.1905\pm 0.0061\,\, (3.2\%)\nonumber\\
d_{0.35}^{WP}&\equiv& r_s(z_d)/D_V(z=0.35)=0.1097\pm 0.0036 \,\, (3.3\%) \nonumber\\
r&=&0.337,
\label{eq:WP}
\ea
where $r$ is the normalized correlation coefficient of $d_{0.2}$
and $d_{0.35}$. The inverse covariance matrix of ($d_{0.2}$, $d_{0.35}$) 
is \cite{Percival10}
\be
C_{\rm GC,WP}^{-1}=\left( \begin{array}{cc}
   30124  & -17227\\
   -17227  & 86977
   \end{array}
\right)
\label{eq:WP_cov}
\ee

For comparison, we also consider the distance measurements
from the WiggleZ survey at $z=0.6$ by Blake et al. (2011) \cite{Blake11}, 
and from the 6dF GRS at $z=0.106$ by Beutler et al. (2011) \cite{6dF}:
\ba
\label{eq:Blake2}
d_{0.6}&\equiv& r_s(z_d)/D_V(z=0.6)= 0.0692\pm 0.0033 \nonumber\\
d_{0.106}&\equiv& r_s(z_d)/D_V(z=0.106)= 0.336\pm 0.015
\ea

GC data are included in our analysis by adding
the following term to the $\chi^2$ of a given model
with 
\be
\label{eq:chi2bao}
\chi^2_{GC}=\Delta p_i \left[ {\rm C}^{-1}_{GC}(p_i,p_j)\right]
\Delta p_j,
\hskip .5cm
\Delta p_i= p_i - p_i^{data}.
\ee
For the Chuang \& Wang (2011) GC measurements \cite{CW11},
$p_1=H(z=0.35)r_s(z_d)$ and $p_2=r_s(z_d)/D_A(z=0.35)$.
For the Percival et al. (2010) GC measurements \cite{Percival10},
$p_1=d_{0.2}$ and $p_2=d_{0.35}$. Note that $p_i^{data}$ ($i=1,2$)
are given in Eqs.(\ref{eq:CW2}) and (\ref{eq:WP}), and
the inverse covariance matrices are given in Eqs.(\ref{eq:CW2_cov}) and 
(\ref{eq:WP_cov}). Since the WiggleZ and 6dF surveys give
independent measurements, Eq.(\ref{eq:chi2bao}) is replaced
with $\chi^2_{GC}=\sum_{i=1}^{2} (\Delta p_i)^2 /\sigma_{p_i}^2$,
with $p_1=d_{0.6}$, and $p_2=d_{0.106}$, and $\sigma_{p_i}$ 
($i=1,2$) given in Eq.(\ref{eq:Blake2}).

\subsection{CMB data}
\label{sec:CMB}

CMB data give us the comoving distance to the photon-decoupling surface 
$r(z_*)$, and the comoving sound horizon 
at photo-decoupling epoch $r_s(z_*)$ \cite{Page03}.
Wang \& Mukherjee 2007 \cite{WangPia07} showed that
the CMB shift parameters
\ba
R &\equiv &\sqrt{\Omega_m H_0^2} \,r(z_*)/c, \nonumber\\
l_a &\equiv &\pi r(z_*)/r_s(z_*),
\ea
together with $\omega_b\equiv \Omega_b h^2$, provide an efficient summary
of CMB data as far as dark energy constraints go.
This has been verified by \cite{Li08}.
Replacing $\omega_b$ with $z_*$ gives identical
constraints when the CMB distance priors are
combined with other data \cite{Wang08b}. 

The comoving sound horizon at redshift $z$ is given by
\ba
\label{eq:rs}
r_s(z)  &= & \int_0^{t} \frac{c_s\, dt'}{a}
=cH_0^{-1}\int_{z}^{\infty} dz'\,
\frac{c_s}{E(z')}, \nonumber\\
 &= & cH_0^{-1} \int_0^{a} 
\frac{da'}{\sqrt{ 3(1+ \overline{R_b}\,a')\, {a'}^4 E^2(z')}},
\ea
where $a$ is the cosmic scale factor, $a =1/(1+z)$, and
$a^4 E^2(z)=\Omega_m (a+a_{\rm eq})+\Omega_k a^2 +\Omega_X X(z) a^4$,
with $a_{\rm eq}=\Omega_{\rm rad}/\Omega_m=1/(1+z_{\rm eq})$, and
$z_{\rm eq}=2.5\times 10^4 \Omega_m h^2 (T_{CMB}/2.7\,{\rm K})^{-4}$.
The sound speed is $c_s=1/\sqrt{3(1+\overline{R_b}\,a)}$,
with $\overline{R_b}\,a=3\rho_b/(4\rho_\gamma)$,
$\overline{R_b}=31500\Omega_bh^2(T_{CMB}/2.7\,{\rm K})^{-4}$.
We take $T_{CMB}=2.725$.

The redshift to the photon-decoupling surface, $z_*$, is given by the 
fitting formula \cite{Hu96}:
\be
z_*=1048\, \left[1+ 0.00124 (\Omega_b h^2)^{-0.738}\right]\,
\left[1+g_1 (\Omega_m h^2)^{g_2} \right],
\ee
where
\ba
g_1 &= &\frac{0.0783\, (\Omega_b h^2)^{-0.238}}
{1+39.5\, (\Omega_b h^2)^{0.763}}\\
g_2 &= &\frac{0.560}{1+21.1\, (\Omega_b h^2)^{1.81}}
\ea
The redshift of the drag epoch $z_d$ is well approximated by 
\cite{EisenHu98}
\begin{equation}
z_d  =
 \frac{1291(\Omega_mh^2)^{0.251}}{1+0.659(\Omega_mh^2)^{0.828}}
\left[1+b_1(\Omega_bh^2)^{b2}\right],
\label{eq:zd}
\end{equation}
where
\begin{eqnarray}
  b_1 &= &0.313(\Omega_mh^2)^{-0.419}\left[1+0.607(\Omega_mh^2)^{0.674}\right],\\
  b_2 &= &0.238(\Omega_mh^2)^{0.223}.
\end{eqnarray}

%\begin{figure} 
%\psfig{file=la.ps,width=3in}\\
%\vspace{-0.4in}
%\caption{\label{fig:l_a}\footnotesize%
%One-dimensional marginalized probability distributions
%of CMB shift parameter $l_a$ derived from WMAP7 data.
%
%\end{figure}

%\begin{figure} 
%\psfig{file=R.ps,width=3in}\\
%\vspace{-0.4in}
%\caption{\label{fig:R}\footnotesize%
%One-dimensional marginalized probability distribution 
%of CMB shift parameter $R$ derived from WMAP7 data.
%}
%\end{figure}

%\begin{figure} 
%\psfig{file=zstar.ps,width=3in}\\
%\vspace{-0.4in}
%\caption{\label{fig:zstar}\footnotesize%
%One-dimensional marginalized probability distribution 
%of the redshift to the photon-decoupling surface, $z_*$, 
%derived from WMAP7 data.
%}
%\end{figure}

%\begin{figure} 
%\psfig{file=ns.ps,width=3in}\\
%\vspace{-0.4in}
%\caption{\label{fig:ns}\footnotesize%
%One-dimensional marginalized probability distributions
%of powerlaw index of primordial matter power spectrum, $n_s$,
%derived from WMAP7 data.
%}
%\end{figure}

%\begin{figure} 
%\psfig{file=rs.ps,width=3in}\\
%\vspace{-0.4in}
%\caption{\label{fig:rs(zd)}\footnotesize%
%One-dimensional marginalized probability distribution
%of the sound horizon at the drag epoch, $r_s(z_d)$,
%derived from WMAP7 data.
%}
%\end{figure}

%Figs.\ref{fig:l_a}-\ref{fig:rs(zd)} show the 
We have derived the one-dimensional marginalized probability 
distributions (pdf) of $(l_a, R, z_*, n_s, r_s(z_d))$ from WMAP7 data,
for three different assumptions about dark energy:
(1) $w_X(z)=-1$; (2) $w_X(z)=w$ (constant);
(3) $w_X(z)=w_0+w_a(1-a)$ \cite{Chev01}.
We find that these pdf's are nearly independent of the
assumption about dark energy, and are well fitted by Gaussian
distributions with the following means and standard deviations:
\ba
&&\langle l_a \rangle = 302.35, \sigma(l_a)=0.86 \nonumber\\
&&\langle R \rangle = 1.728,  \sigma(R)=0.02\nonumber\\
&& \langle z_* \rangle = 1091.32, \sigma(z_*)=1.0\nonumber\\
&& \langle n_s \rangle = 0.963, \sigma(n_s)=0.016  \nonumber\\
&&  \langle r_s(z_d) \rangle = 152.85\,{\rm Mpc}, \sigma(r_s[z_d])=1.85\,{\rm Mpc}.
\label{eq:CMB_mean}
\ea
%The arrows in Figs.\ref{fig:l_a}-\ref{fig:rs(zd)} indicate the
Our Gaussian fits are fully consistent with the bestfit values for 
a constant $w$ from Komatsu's website \cite{Komatsu10}.

The normalized covariance matrix of $(l_a, R, z_*, n_s, r_s(z_d))$ is
\be
\left(
\begin{array}{ccccc}
  1.0000 &  .20794 &   .47422  & -.54889 &  .34914\\
   .20794 & 1.0000 &   .74409  & -.44370 & -.76929\\
   .47422 &  .74409 &  1.0000  & -.79575 & -.19121\\
  -.54889 & -.44370 &  -.79575  & 1.0000 & -.09973\\
   .34914 & -.76929 &  -.19121  & -.09973 & 1.0000
\end{array}
\right)
\ee

We find that the addition of $r_s(z_d)$ is redundant, and amounts
to partially double-weighting the CMB constraints;
so its measurement cannot be used in the CMB priors 
unless it is used to replace $l_a(z_*)$. 
This was not apparent when WMAP5 data were used \cite{Wang09}; 
this is because WMAP7 data provide significantly tighter constraints on the
set of CMB distance priors.

Since the primary GC data we use in this paper have been marginalized 
over $n_s$ \cite{CW11}, we should marginalized the CMB distance priors 
over $n_s$ as well. When marginalized over $n_s$ and $r_s(z_d)$, the 
inverse covariance matrix of $(l_a, R, z_*)$ from WMAP7 is
\be
\mbox{Cov}_{CMB}^{-1}=
\left(
\begin{array}{ccc}
   1.8571  & 25.929 & -1.1433\\
   25.929  & 5963.3 & -99.319\\
  -1.1433  &-99.319  & 2.9443
\label{eq:CMB_cov}
\end{array}
\right)
\ee

CMB data are included in our analysis by adding
the following term to the $\chi^2$ of a given model
with  $p_1=l_a(z_*)$, $p_2=R(z_*)$,and $p_3=z_*$:
\be
\label{eq:chi2CMB}
\chi^2_{CMB}=\Delta p_i \left[ \mbox{Cov}^{-1}_{CMB}(p_i,p_j)\right]
\Delta p_j,
\hskip .5cm
\Delta p_i= p_i - p_i^{data},
\ee
where $p_i^{data}$ are the mean from Eq.(\ref{eq:CMB_mean}),
and ${\rm Cov}^{-1}_{CMB}$ is the inverse covariance matrix of 
[$R(z_*), l_a(z_*), z_*]$ from Eq.(\ref{eq:CMB_cov}).
Note that $p_4=n_s$ should be added if the constraints on
$n_s$ are included in the GC data.
Finally, our Gaussian fit for the pdf of $r_s(z_d)$ should
be useful when fitting for BAO peak locations.

\subsection{Gammay-ray Burst Data}
\label{sec:GRB}

We add gammay-ray burst (GRB) data to our analysis, since these are
complementary in redshift range to the SN Ia data.
We use GRB data in the form of the model-independent GRB distance 
measurements from Wang (2008c) \cite{Wang08c}, which were
derived from the data of 69 GRBs with $0.17 \le z \le 6.6$
from Schaefer (2007) \cite{Schaefer07}\footnote{The proper
calibration of GRBs is an active area of research. For 
recent studies on the impact of detector thresholds, 
spectral analysis, and unknown selection effects, see, e.g., 
\cite{Butler07,Butler09,Petrosian09,Shahmoradi11}.}.

The GRB distance measurements are given in terms of \cite{Wang08c}
\be
\label{eq:rp}
\overline{r_p}(z_i)\equiv \frac{r_p(z)}{r_p(0.17)}, \hskip 1cm
r_p(z) \equiv \frac{(1+z)^{1/2}}{z}\, \frac{H_0}{ch}\, r(z),
\ee
where $r(z)$ is the comoving distance at $z$.

The GRB data are included in our analysis by adding
the following term to the $\chi^2$ of a given model:
\ba
\label{eq:rGRB1}
\chi^2_{GRB} &= & \left[\Delta \overline{r_p}(z_i)\right]  \cdot
\left(\mathrm{Cov}^{-1}_{GRB}\right)_{ij}\cdot
\left[\Delta \overline{r_p}(z_j)\right]
\nonumber\\
\Delta \overline{r_p}(z_i) &= & \overline{r_p}^{\mathrm{data}}(z_i)-\overline{r_p}(z_i),
\ea
where $\overline{r_p}(z)$ is defined by Eq.(\ref{eq:rp}).
The covariance matrix is given by
\be
\left(\mathrm{Cov}_{GRB}\right)_{ij}=
\sigma(\overline{r_p}(z_i)) \sigma(\overline{r_p}(z_j)) 
\left(\overline{\mathrm{Cov}}_{GRB}\right)_{ij},
\ee
where $\overline{\mathrm{Cov}}_{GRB}$ is the normalized covariance matrix
from Table 3 of Wang (2008c) \cite{Wang08c}, and
\ba
\label{eq:rGRB3}
\sigma(\overline{r_p}(z_i))  &= &\sigma\left(\overline{r_p}(z_i)\right)^+, \hskip 0.5cm \mathrm{if}\,\, 
\overline{r_p}(z) \ge \overline{r_p}(z)^{\mathrm{data}}; \nonumber\\
\sigma(\overline{r_p}(z_i))  &= &\sigma\left(\overline{r_p}(z_i)\right)^-, \hskip 0.5cm \mathrm{if}\,\, 
\overline{r_p}(z) < \overline{r_p}(z)^{\mathrm{data}},
\ea
where $\sigma\left(\overline{r_p}(z_i)\right)^+$ and 
$\sigma\left(\overline{r_p}(z_i)\right)^-$ are the 68\% C.L. errors
given in Table 2 of Wang (2008c) \cite{Wang08c}.

\subsection{Dark energy parametrization}
\label{sec:para}

Since we are ignorant of the true nature of dark energy,
it is useful to measure the dark energy density function
$X(z)\equiv \rho_X(z)/\rho_X(0)$ as a free function of
redshift \cite{WangGarnavich,WangTegmark04,WangFreese06}.
This has the advantage of allowing dark energy
models in which $\rho_X(z)$ becomes negative in the future,
e.g., the ``Big Crunch'' models \cite{Linde87,WangLinde04},
which are precluded if we parametrize dark energy with 
an equation of state $w_X(z)$ \cite{WangTegmark04}.

Here we parametrize $X(z)$ by cubic-splining its values
at $z=1/3$, $2/3$, and 1.0, and assume that
$X(z>1)=X(z=1)$. For simplicity of notation,
we define $X_{0.33}\equiv X(z=1/3)$,
$X_{0.67}\equiv X(z=2/3)$, and $X_{1.0}\equiv X(z=1)$.
Fixing $X(z>1)$ reflects the limit
of current data, and avoids making assumptions about early
dark energy that can be propagated into artificial constraints on
dark energy at low $z$ \cite{WangTegmark04,WangPia07}.

For comparison with the work of others, we also 
consider a dark energy equation of state linear in the cosmic scale 
factor $a$, $w_X(a)=w_0+(1-a)w_a$ \cite{Chev01}.
A related parametrization is \cite{Wang08b}
\be
w_X(z)=w_0(3a-2)+ 3w_{0.5}(1-a),
\ee
where $w_{0.5}\equiv w_X(z=0.5)$.
Wang (2008b) \cite{Wang08b} showed that ($w_0$, $w_{0.5}$) are much less
correlated than ($w_0$, $w_a$), thus are a better set of parameters
to use. We find that ($w_0$, $w_{0.5}$) converge much faster
than ($w_0$, $w_a$) in a Markov Chain Monte Carlo (MCMC)
likelihood analysis for the same data.

\section{Results}

We perform a MCMC likelihood analysis \cite{Lewis02} to obtain 
${\cal O}$($10^6$) samples for each set of results presented in 
this paper. We assume flat priors for all the parameters, and allow ranges 
of the parameters wide enough such that further increasing the allowed 
ranges has no impact on the results.
The chains typically have worst e-values, defined to be
the variance(mean)/mean(variance) of 1/2 chains,
much smaller than 0.005, indicating convergence.
The chains are subsequently appropriately thinned to 
ensure independent samples.

In addition to the SN Ia, CMB, GC, and GRB data discussed in 
Sec.{\ref{sec:method}}, we impose a prior of 
$H_0 = 73.8 \pm 2.4\,$km$\,$s$^{-1}$Mpc$^{-1}$, from the
HST measurements by Riess et al. (2011) \cite{Riess11}.

We do {\it not} assume a flat universe unless specifically noted.
In addition to the dark energy parameters described in Sec.\ref{sec:para},
we also constrain cosmological parameters ($\Omega_m, \Omega_k, h, \omega_b$), 
where $\omega_b\equiv \Omega_b h^2$. In addition, we marginalize over
the SN Ia nuisance parameters $\{\alpha, \beta, {\cal M}\}$.

We will present results for dark energy density at
$z=1/3$, $2/3$, and 1, as well as ($w_0,w_a$) and 
($w_0,w_{0.5}$),
and a constant dark energy equation of state $w$.

\subsection{The effect of flux-averaging of SNe Ia}

\begin{figure} 
\psfig{file=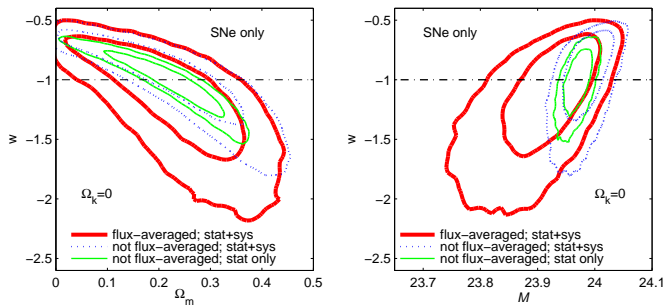,width=3.5in}\\
%\vspace{-0.4in}
\caption{\label{fig:sn_flux}\footnotesize%
The 2D marginalized contours of $(w, \Omega_m, {\cal M})$
for SNe data (with and without flux-averaging), assuming a flat universe.
The contours are at 68\% and 95\% confidence levels.
}
\end{figure}

Figs.\ref{fig:sn_flux} shows the 2D marginalized contours of
$(w, \Omega_m, {\cal M})$, assuming a constant
equation of state for dark energy, $w$, and a flat universe.
Note that the inclusion of systematic errors of SNe 
leads to significantly larger uncertainties in
estimated parameters, compared to
when only statistical errors of SNe are included.
Clearly, flux-averaging (thick solid lines) leads to 
larger errors on dark energy and 
cosmological parameters if only SN Ia data are used. 

\begin{figure} 
\psfig{file=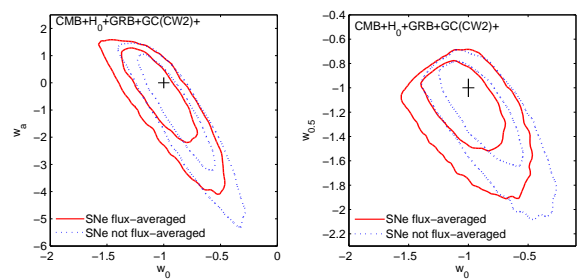,width=3in}\\
%\vspace{-0.4in}
\caption{\label{fig:w0w1_flux_GC}\footnotesize%
The 2D marginalized contours of $(w_0,w_a)$ and $(w_0,w_{0.5})$
for SNe data (with and without flux-averaging) combined with galaxy 
clustering (CW2), CMB, $H_0$, and GRB data.
The contours are at 68\% and 95\% confidence levels.
}
\end{figure}

Fig.\ref{fig:w0w1_flux_GC} shows the 2D marginalized contours of 
$(w_0,w_a)$ and $(w_0,w_{0.5})$ for SNe data combined with 
with CMB, $H_0$, GRB, and GC (CW2) data.
The solid contours are for flux-averaged SNe,
and dotted contours are for SNe without flux-averaging.
The SNe data with or without flux-averaging lead to
qualitatively consistent results, with flux-averaging
expanding the parameter space at $w_0<-1$, which
shifts the bestfit model from $w_0>-1$ towards $w_0=-1$,
and from $w_a<0$ towards $w_a=0$
(it has a smaller impact on $w_{0.5}$ since $w_{0.5}$ is
less correlated with $w_0$ than $w_a$).
Table \ref{table:w0w1} tabulates the mean and 68.3\%
confidence intervals for $(w_0,w_a)$ and $(w_0,w_{0.5})$
for SNe data combined with with CMB, $H_0$, GRB, 
with or without GC (CW2) data (the latter
corresponds to Fig.\ref{fig:w0w1_flux_GC}).

\begin{table*}[htb]
\caption{Effect of flux-averaging SNe on $(w_0,w_a)$ and $(w_0,w_{0.5})$}
\label{table:w0w1}
\begin{center}
\begin{tabular}{lcccc}
\hline
\hline
flux-averaging & $w_0$  &  $w_a$  & $w_0$ & $w_{0.5}$ \\
\hline
& SNe+CMB+$H_0$+GRB  & & & \\
yes & $-0.987$  ($-1.247, -0.727$)\, & $-0.780$  ($-2.046, 0.516$) \,
    & $-0.997$ ($-1.268, -0.722$)\,& $-1.260$ ($-1.546, -0.965$) \\
no &  $-0.780$  ($-1.013, -0.545$) & $-1.424$  ($-2.891, 0.045$) 
   &  $-0.783$ ($-1.023, -0.547$) & $-1.248$ ($-1.549, -0.946$) \\
\hline  
& SNe+CMB+$H_0$+GRB+GC(CW2)  & & & \\
yes &     $-0.995$ ($-1.211, -0.776$) & $-0.676$ ($-1.884, 0.566)$ 
&    $-1.000$  ($-1.216, -0.784)$ & $-1.211$  ($-1.454, -0.966)$  \\
no &     $-0.832$  ($-1.061, -0.606)$ & $-1.353$ ($-2.844,  0.151)$ 
&    $-0.842$  ($-1.067, -0.615)$ & $-1.266$  ($-1.559, -0.972)$ \\
\hline
\end{tabular}
\end{center}
\end{table*}
    
%\begin{figure} 
%\psfig{file=snls3wmap7yr3norsHSTR11GRB_w0w1_flux_comp_2D.ps,width=3in}\\
%\caption{\label{fig:w0w1_flux_noGC}\footnotesize%
%The 2D marginalized contours of $(w_0,w_a)$ and $(w_0,w_{0.5})$
%for SNe data (with and without flux-averaging) combined with CMB, $H_0$, and GRB data.
%The contours are at 68\% and 95\% confidence levels.
%}
%\end{figure}

\begin{figure} 
\psfig{file=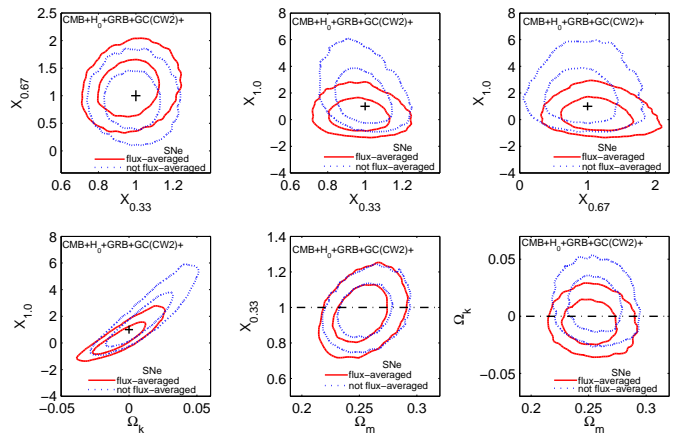,,width=3.5in}\\
%\vspace{-0.4in}
\caption{\label{fig:X3_flux_GC}\footnotesize%
The 2D marginalized contours of $(X_{0.33}, X_{0.67}, X_{1.0}, \Omega_m, \Omega_k)$
for SNe data (with and without flux-averaging) combined with galaxy clustering (CW2), 
CMB, $H_0$, and GRB data.
The contours are at 68\% and 95\% confidence levels.
}
\end{figure}

Fig.\ref{fig:X3_flux_GC} shows the 2D marginalized contours 
of $(X_{0.33}, X_{0.67}, X_{1.0}, \Omega_m, \Omega_k)$
for SNe data combined with CMB, $H_0$, GRB,
and GC data (CW2). The solid contours are for flux-averaged SNe,
and dotted contours are for SNe without flux-averaging.
The SNe data with or without flux-averaging lead to
qualitatively consistent results, with a shift of the
bestfit model towards a cosmological constant; this is
the same trend as in the $(w_0,w_a)$ and $(w_0,w_{0.5})$ parametrizations.
Note that flux-averaging leads to significantly tighter constraints on 
the dark energy density at $z=1$, $X_{1.0}$; this indicates
that the impact of flux-averaging increases with redshift,
since $(X_{0.33}, X_{0.67}, X_{1.0})$ are only weakly correlated.
This is as expected, since the bias in estimated parameters
due to weak lensing increases with redshift \cite{Wang05}.

%\begin{figure} 
%\psfig{file=snls3wmap7yr3norsHSTR11GRB_fluxavg_X3zcut1flat_comp_2D.ps,width=3.5in}\\
%\caption{\label{fig:X3_flux_noGC}\footnotesize%
%The 2D marginalized contours of $(X_{0.33}, X_{0.67}, X_{1.0}, \Omega_m, \Omega_k)$
%for SNe data (with and without flux-averaging) combined with CMB, $H_0$, and GRB data.
%The contours are at 68\% and 95\% confidence levels.
%}
%\end{figure}

\subsection{Comparison of different galaxy clustering data}

\begin{figure} 
\psfig{file=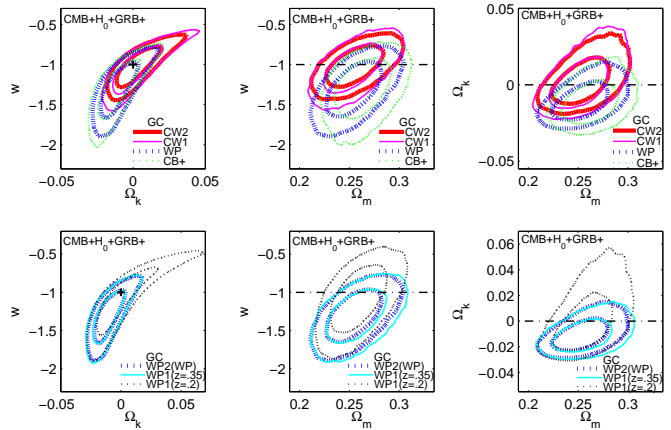,,width=3.5in}\\
%\vspace{-0.4in}
\caption{\label{fig:w_GC}\footnotesize%
The 2D marginalized contours of $(w, \Omega_m, \Omega_k)$
for different galaxy clustering measurements combined with 
CMB, $H_0$, and GRB data.
The contours are at 68\% and 95\% confidence levels.
}
\end{figure}

We now explore the differences of the various galaxy clustering
(GC) measurements in constraining dark energy,
assuming a constant equation of state for dark energy, $w$.
Fig.\ref{fig:w_GC} shows the 2D marginalized contours of $(w, \Omega_m, \Omega_k)$
for different GC measurements combined with 
CMB, $H_0$, and GRB data.

The first row of Fig.\ref{fig:w_GC} compares the
$H(z=0.35)r_s(z_d)$ and $r_s(z_d)/D_A(z=0.35)$ measurements by 
Chuang \& Wang (2011) \cite{CW11} with their $d_{0.35}=r_s(z_d)/D_V(z=0.35)$ 
measurement (both from SDSS DR7 LRGs), as well as the
$d_{0.2}$ and $d_{0.35}$ measurements by Percival et al. (2010) 
\cite{Percival10} from SDSS DR7 LRG and main galaxy samples and 2dFGRS,
and the $d_{0.6}$ measurement by Blake et al. (2011) from
the WiggleZ survey \cite{Blake11} combined with the
$d_{0.106}$ measurement by Beutler et al. (2011) from 6dF GRS.

For the Chuang \& Wang (2011) \cite{CW11} GC measurements (CW2
and CW1), the constraints on $w$ are tightened significantly by going 
from spherically-averaged data (CW1), i.e., $d_{0.35}$, to 2D data (CW2), i.e.,
$H(z=0.35)r_s(z_d)$ and $r_s(z_d)/D_A(z=0.35)$, as indicated
by comparing the thin solid contours (CW1) to thick solid contours 
(CW2) in the first row of Fig.\ref{fig:w_GC}.
This is as expected, as more information from GC is included in CW2
compared to CW1.

Both the Percival et al. (2010) GC measurements (WP) and the
combined WiggleZ survey and 6dF GRS measurements (CB+)
favor $w<-1$, while the Chuang \& Wang (2011) \cite{CW11} GC measurements
favor $w=-1$.

The second row in Fig.\ref{fig:w_GC} compares the 
$d_{0.2}$ and $d_{0.35}$ measurements by Percival et al. (2010) 
\cite{Percival10} (WP2), with their measurements of
$d_{0.2}$ and $d_{0.35}$ separately.
Clearly, most of the constraining power on $w$ comes from 
$d_{0.35}$. While the $d_{0.2}$ measurement favors
$w=-1$, the $d_{0.35}$ measurement favors
$w<-1$.

The measurements of $d_{0.35}$ by Chuang \& Wang (2011) \cite{CW11}
and Percival et al. (2010) \cite{Percival10} are similar
in precision, but differ systematically:
$d_{0.35}^{CW} \equiv r_s(z_d)/D_V(z=0.35)=0.1161\pm   0.0034$,
while $d_{0.35}^{WP}\equiv r_s(z_d)/D_V(z=0.35)=0.1097\pm 0.0036$.
The lower measured value of $d_{0.35}^{WP}$ implies 
a smaller $H(z=0.35)$, which in turn implies a more
negative $w$. When combined with CMB, $H_0$, and GRB data,
$d_{0.35}^{CW}$ favors $w=-1$, while $d_{0.35}^{WP}$ favors 
$w<-1$. Note that these two measurements used different
methods to analyze GC data: Chuang \& Wang (2011) used
the galaxy correlation function, while Percival et al. (2010)
used galaxy power spectrum. It is not surprising that
they lead to different distance measurements from GC.

For the rest of this paper, we will present only results using 
the GC measurements by Chuang \& Wang (2011) \cite{CW11}; these
are more conservative as they were derived without assuming 
CMB priors.

\subsection{Constraints on dark energy and $H(z)$}

\begin{figure} 
\psfig{file=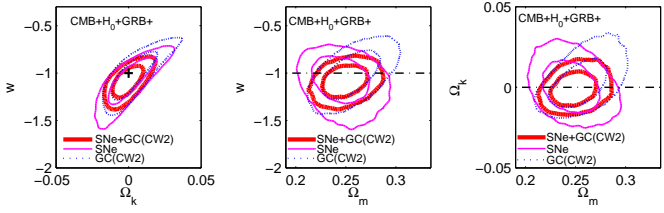,width=3.5in}\\
%\vspace{-0.4in}
\caption{\label{fig:w_all}\footnotesize%
The 2D marginalized contours of $(w, \Omega_m, \Omega_k)$
for SNe data (flux-averaged), galaxy clustering measurements (CW2),
combined with CMB, $H_0$, and GRB data.
The contours are at 68\% and 95\% confidence levels.
}
\end{figure}

\begin{figure} 
\psfig{file=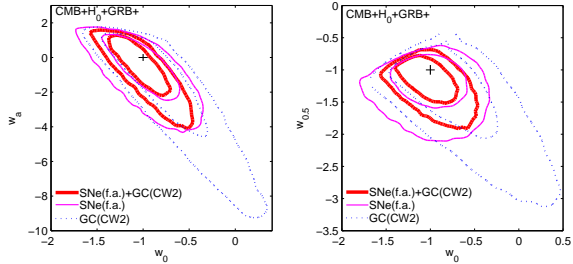,width=3in}\\
%\vspace{-0.4in}
\caption{\label{fig:w0w1_all}\footnotesize%
The 2D marginalized contours of $(w_0,w_a)$ and $(w_0,w_{0.5})$
for SNe data (flux-averaged), galaxy clustering measurements (CW2),
combined with CMB, $H_0$, and GRB data.
The contours are at 68\% and 95\% confidence levels.
}
\end{figure}

\begin{figure} 
\psfig{file=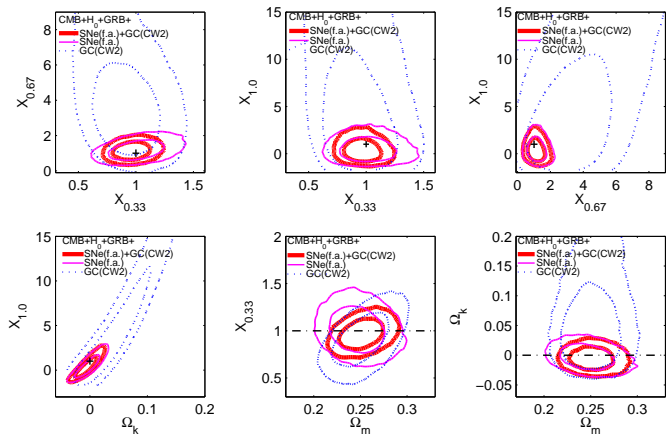,width=3.5in}\\
%\vspace{-0.4in}
\caption{\label{fig:X3_all}\footnotesize%
The 2D marginalized contours of $(X_{0.33}, X_{0.67}, X_{1.0}, \Omega_m, \Omega_k)$
for SNe data (flux-averaged), galaxy clustering measurements (CW2),
combined with CMB, $H_0$, and GRB data.
The contours are at 68\% and 95\% confidence levels.
}
\end{figure}

Figs.\ref{fig:w_all}-\ref{fig:X3_all} show the
2D marginalized contours of dark energy and cosmological
parameters for the three different sets of dark energy
parameters: (1) $w_X(z)=w$=const.; 
(2) $w_X(z)=w_0+w_a(1-a)$, and 
$w_X(z)=w_0(3a-2)+ 3w_{0.5}(1-a)$;
(3) $X(z)\equiv \rho_X(z)/\rho_X(0)$ given by
a cubic spline at $z<1$ over $X_{z_i}\equiv X(z_i)$, with
$z_1=1/3$, $z_2=2/3$, and $z_3=1$, and $X(z>1)=X(z=1)$.

\begin{figure} 
\psfig{file=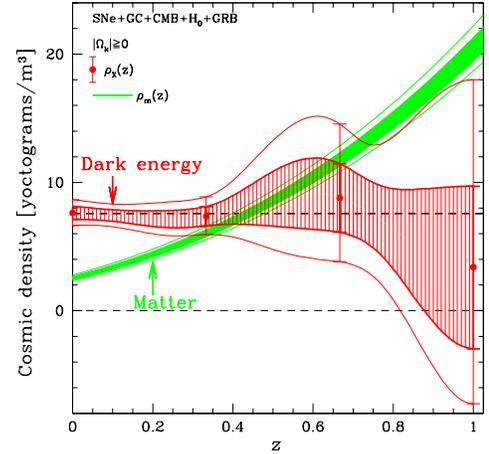,width=3in}\\
\psfig{file=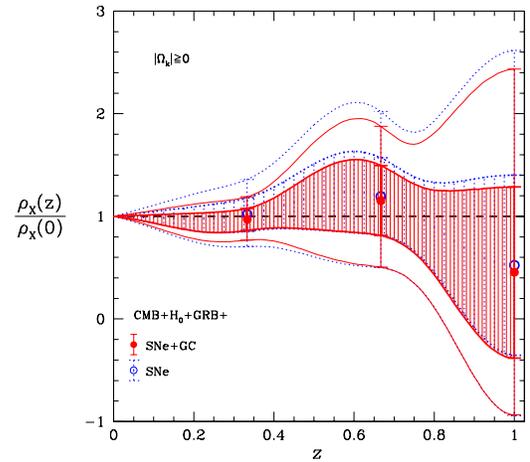,width=3in}\\
%\vspace{-0.4in}
\caption{\label{fig:rhoX}\footnotesize%
Dark energy density function $\rho_X(z)$ measured from
SNe data (flux-averaged with $dz=0.07$), galaxy clustering measurements (CW2),
combined with CMB, $H_0$, and GRB data.
The contours are at 68\% and 95\% confidence levels.
The upper panel shows both $\rho_X(z)$ and $\rho_m(z)$
in units of yoctograms per cubic meter, with
1 yoctogram $= 10^{-24}$ grams. The lower panel 
shows the corresponding $X(z)=\rho_X(z)/\rho_X(0)$.
}
\end{figure}

\begin{figure} 
\psfig{file=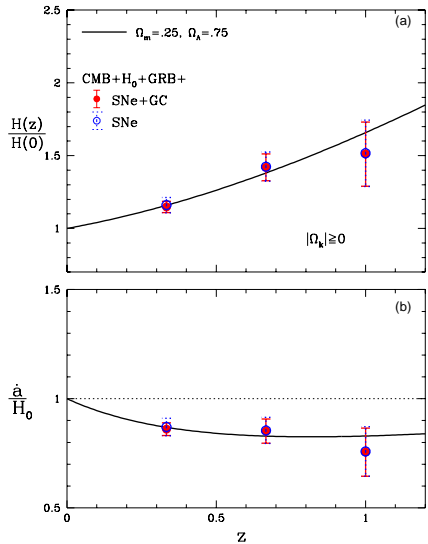,width=3in}\\
%\vspace{-0.4in}
\caption{\label{fig:Hz}\footnotesize%
The cosmic expansion history $H(z)$ measured from
SNe data (flux-averaged with $dz=0.07$), galaxy clustering measurements (CW2),
combined with CMB, $H_0$, and GRB data. The error bars indicate
the 68.3\% confidence level ranges.
}
\end{figure}

Fig.\ref{fig:rhoX} shows the dark energy density function $\rho_X(z)$ 
measured from SNe data (flux-averaged with $dz=0.07$), galaxy clustering 
measurements (CW2), combined with CMB, $H_0$, and GRB data.
The upper panel of Fig.\ref{fig:rhoX} shows both $\rho_X(z)$
and $\rho_m(z)$, in units of yoctograms per cubic meter, with
1 yoctogram $= 10^{-24}$ grams. The uncertainties in
$X(z)$, $\Omega_X=1-\Omega_m-\Omega_k$, and $h$ have been propagated into
that of $\rho_X(z)$. The lower panel of Fig.\ref{fig:rhoX}
shows the corresponding $X(z)=\rho_X(z)/\rho_X(0)$.
Fig.\ref{fig:Hz} shows the corresponding cosmic expansion history $H(z)$.

Clearly, given the same priors of CMB, $H_0$, and GRB data,
SNe lead to much stronger constraints on dark energy than
galaxy clustering data at present. Note also that the addition 
of galaxy clustering data to SN data leads to significantly
improved constraints.

Table \ref{table:Xz,Hz} gives the dark energy density function 
$X(z)\equiv \rho_X(z)/\rho_X(0)$ and the cosmic expansion history 
$H(z)$, as well as ($\Omega_m, \Omega_k, h, \omega_b)$,
measured from current data of SNe+GC+CMB+$H_0$+GRB. 
The $H(z)$ measurements are derived using
Eq.(\ref{eq:H(z)}). Tables \ref{table:Xz_cov}-\ref{table:Hz_cov} 
give the normalized covariance matrices of the $X(z)$ and $H(z)$ 
measurements. Note that both the $X(z)$ and $H(z)$ measurements are 
only weakly correlated at different redshifts.

\begin{table*}[htb]
\caption{$X(z)$, $H(z)$, and cosmological parameters
estimated from current data with 68.3\% C.L. upper and lower uncertainties.}
\label{table:Xz,Hz}
\begin{center}
\begin{tabular}{|l|llcc|}
\hline 
 & $\mu$ & $\sigma$ & $\mu+\sigma^{-}$ & $\mu-\sigma^+$ \\
 \hline 
$X(z=1/3)$ &  0.969 &  0.108  &0.862  &1.077  \\   
$ X(z=2/3)$ &  1.152 & 0.347 & 0.812 & 1.492\\   
$ X(z=1.0)$ &  0.453 & 0.863 &$-0.385$ & 1.284  \\  
$\Omega_m$ & 0.252& 0.016 & 0.236 & 0.268   \\    
$\Omega_k$ & $-0.0049$ & 0.0131& $-0.0180 $& 0.0080  \\   
$h$        &  0.734 & 0.019 & 0.715 & 0.753  \\    
$\omega_b$ &  0.02234 & 0.00062 & 0.02170 & 0.02290  \\   
\hline
$H(z=1/3)/H_0$ &  1.148 & 0.040&  1.107 & 1.186  \\   
$H(z=2/3)/H_0$ &  1.419 & 0.093&  1.328 & 1.510  \\    
$H(z=1.0)/H_0$ &  1.511 & 0.221&  1.289 & 1.731  \\    
\hline
\end{tabular}
%\tablecomments{}
\end{center}
\end{table*}

\begin{table*}[htb]
\caption{Normalized covariance matrix of $X(z)$ from current data}
\label{table:Xz_cov}
\begin{center}
\begin{tabular}{|l|rrr|}
\hline
       & 1  & 2  & 3  \\
    \hline
1&  1.0000 & 0.1830 &$ -0.1195$\\
2 & 0.1830 & 1.0000 &$ -0.1283$\\
3 &$ -0.1195$ &$ -0.1283$ & 1.0000\\
\hline
\end{tabular}
%\tablecomments{}
\end{center}
\end{table*}

\begin{table*}[htb]
\caption{Normalized covariance matrix of $H(z)/H_0$ from current data}
\label{table:Hz_cov}
\begin{center}
\begin{tabular}{|l|rrr|}
\hline
       & 1  & 2  & 3    \\
    \hline
1&     1.0000 & 0.2657 & 0.0727\\
2 &  0.2657 & 1.0000 &$-0.0575$\\
3  &0.0727 &$-0.0575$ & 1.0000\\
\hline
\end{tabular}
%\tablecomments{}
\end{center}
\end{table*}

To quantify the comparison in dark energy constraints, we 
can use the general dark energy Figure-of-Merit (FoM)
defined by Wang (2008b) \cite{Wang08b} 
\be
{\rm FoM} = \frac{1}{\sqrt{{\rm det Cov}(f_1, f_2, ...,f_N)}}
\label{eq:FoM}
\ee
where $(f_1, f_2, ...,f_N)$ is the set of parameters that have been
chosen to parametrize dark energy. 
The Dark Energy Task Force (DETF) defined the dark energy FoM to be 
the inverse of the area enclosed by the 95\% confidence level contour 
of $(w_0,w_a)$ \cite{detf}. The areas enclosed by contours are difficult 
to calculate for real data, as these contours can be quite irregular.
The definition of Eq.(\ref{eq:FoM}) has the advantage of being
easy to calculate for either real or simulated data.
For $(f_1, f_2)=(w_0,w_a)$, ${\rm FoM}$ of Eq.(\ref{eq:FoM})
is proportional to the FoM defined by the DETF for ideal Gaussian-distributed
data, and the same as the relative FoM used by the DETF in
Fisher matrix forecasts.

Table \ref{table:FoM} shows the dark energy FoM from SNe (flux-averaged), 
galaxy clustering measurements (CW2), together with CMB, $H_0$, and GRB data, for 
$(w_0,w_a)$, $(w_0,w_{0.5})$, and $(X_{0.33}, X_{0.67}, X_{1.0})$.

\begin{table*}[htb]
\caption{Dark energy FoM from current data}
\label{table:FoM}
\begin{center}
\begin{tabular}{|l|c|cccc|cccc|}
\hline
CMB+$H_0$+GRB+ & FoM$_r(\{X_i\})$ & $\sigma(w_0)$ & $\sigma(w_{0.5})$ & 
$r_{w_0,w_{0.5}}$ & FoM$_r(w_0,w_{0.5})$ &
$\sigma(w_0)$ & $\sigma(w_a)$ & $r_{w_0,w_a}$ &FoM$_r(w_0,w_a)$\\
\hline
SNe  & 18.61 & 0.27  & 0.33 & $-0.12$ & 11.02 & 0.27  & 1.32  & $-0.70$ & 3.98 \\
GC  & 0.44 & 0.44 & 0.59 & $-0.74$ & 5.76 & 0.39 & 2.40 & $-0.89$ & 2.31 \\
SNe+GC  & 32.00 & 0.22 & 0.27  & $-0.51$ & 19.80  &0.22   & 1.24  & $-0.84$ & 6.74 \\
\hline
\end{tabular}
\end{center}
\end{table*}

\section{Summary and Discussion}

We have examined how dark energy constraints from current observational data 
depend on the analysis methods used: the analysis of
Type Ia supernovae (SNe Ia), and that of galaxy clustering data (GC).

We generalize the flux-averaging analysis method of SNe Ia to 
allow correlated errors of SNe Ia, in order to reduce the systematic bias 
due to weak lensing of SNe Ia. We find that flux-averaging leads to 
larger errors on dark energy and cosmological parameters if only SN Ia 
data are used (see Fig.\ref{fig:sn_flux}). 
When SN Ia data (the latest compilation by the SNLS team)
are combined with WMAP 7 year results, the latest 
Hubble constant ($H_0$) measurement using the HST, and gamma ray burst 
(GRB) data, flux-averaging of SNe Ia increases the concordance of SNe 
Ia with other data, and shifts the bestfit cosmological model notably 
closer to $w=-1$ (see Fig.\ref{fig:w0w1_flux_GC} and Table \ref{table:w0w1}). 
This leads to significantly tighter constraints on 
the dark energy density at $z=1$, and the cosmic curvature $\Omega_k$
(see Fig.\ref{fig:X3_flux_GC}).
We have made our SN flux-averaging code publicly available 
at http://www.nhn.ou.edu/$\sim$wang/SNcode/.

Note that since the flux-averaging of SNe Ia increases the concordance 
of SN Ia with other data, the {\it combined} data with
flux-averaging of SNe Ia gives comparable constraints for
most parameters, and {\it smaller} uncertainties for 
dark energy density at $z=1$, $X_{1.0}$ (most affected by weak lensing), 
and $\Omega_k$ (strongly correlated with $X_{1.0}$),
compared with combined data with no flux-averaging of SNe Ia.
The combination of concordant data tightens the overall constraints,
while that of discordant data may not.

We find that given the same pirors (CMB+$H_0$+GRB),  
both the Percival et al. (2010) GC measurements \cite{Percival10}
(WP: $d_{0.2}$ and $d_{0.35}$) and the combined WiggleZ survey and 6dF 
GRS measurements \cite{Blake11,6dF} (CB+: $d_{0.6}$ and $d_{0.106}$)
favor $w<-1$\footnote{Similar results were found by \cite{Esca11,XLi11} 
using Percival et al. (2010) GC measurements \cite{Percival10}.}, 
while the Chuang \& Wang (2011) GC measurements
\cite{CW11} (CW2: $H(z=0.35)r_s(z_d)$ and $r_s(z_d)/D_A(z=0.35)$) 
favor $w=-1$ (see Fig.\ref{fig:w_GC}).
This could be due to the difference in the analysis methods
used to derive these GC measurements.
We find that the GC measurements by Chuang \& Wang (2011) are consistent with
SN Ia data (which favor $w=-1$), and tighten the dark energy constraints
significantly when combined with SN data
(see Fig.\ref{fig:w_all}-\ref{fig:X3_all}).

For the convenience of future work, we have derived 
Gaussian fits to the probability distributions of a set of
CMB parameters, [$l_a(z_*), R(z_*), z_*, n_s, r_s(z_d)$],
%(see Figs.\ref{fig:l_a}-\ref{fig:rs(zd)}), 
that summarize the WMAP7 data, as well as well their
covariance matrix. Note that while [$l_a(z_*), R(z_*), z_*$] are 
sufficient for summarizing the CMB priors when $n_s$ dependence 
has been marginalized over in GC analysis, [$l_a(z_*), R(z_*), z_*, n_s$]
need to be used when the constraints on $n_s$ from GC
are included (as will be the case for future GC data).
We find that $r_s(z_d)$ gives redundant information,
its measurement cannot be used in the CMB priors unless
it is used to replace $l_a(z_*)$. Also,
our Gaussian fit for the pdf of $r_s(z_d)$ from WMAP7 can
be useful when fitting for BAO peaks in GC data analysis.

We find that current data are fully consistent with a cosmological 
constant and a flat universe, consistent with the latest findings by others (see, e.g., 
\cite{Komatsu11,XLi11,Bean10,Chen11,Dossett11,Holsclaw11,Liang11,SWang11}).
Since the uncertainties remain large, a deviation from a
cosmological constant is far from being ruled out (see Fig.\ref{fig:rhoX}).

At present, SN data provide much stronger constraints than GC data
(see Table \ref{table:FoM}).
This will change when ambitious galaxy surveys are carried out from
space in the future \cite{Cimatti09,Wang10b}. We can expect
dramatic progress in our measurement and understanding of
dark energy in the next decade or so.

\bigskip

{\bf Acknowledgements}
We are grateful to Alex Conley for communicating details of
the SNLS data, to Eiichiro Komatsu for making the covariance matrix 
for [$l_a(z_*), R(z_*), z_*, r_s(z_d), n_s$] he derived 
from WMAP 7 year data available, and to Max Tegmark for
suggesting the presentation of $\rho_X(z)$ in the
upper panel of Fig.\ref{fig:rhoX}. 
We acknowledge the use of CAMB and CosmoMC.
YW and CHC are supported in part by DOE grant DE-FG02-04ER41305.
PM is supported in part by the L'Oreal UK and Ireland Fellowship For 
Women in Science, and by the University of Sussex.

\end{document}